\documentclass{physeauth}
\usepackage{graphicx}
\usepackage{amsmath}
\usepackage{amssymb}
\usepackage{bm}

\begin{document}

\begin{frontmatter}

\title{Aharonov-Bohm oscillations in disordered nanorings with quantum dots:
Effect of electron-electron interactions}

\author[lpi]{Andrew G. Semenov},
\author[fzk,lpi]{Andrei D. Zaikin \thanksref{thank1}}

\address[lpi]{I.E. Tamm Department of Theoretical Physics, P.N.
Lebedev Physics Institute, 119991 Moscow, Russia}
\address[fzk]{Forschungszentrum Karlsruhe, Institut f\"ur
Nanotechnologie, 76021, Karlsruhe, Germany}

\thanks[thank1]{
Corresponding author.
E-mail: Andrei.Zaikin@int.fzk.de}

\begin{abstract}
We investigate the effect of electron-electron interactions on
Aharonov-Bohm (AB) current oscillations in nanorings formed by a
chain of metallic quantum dots.  We demonstrate that
electron-electron interactions cause electron dephasing thereby
suppressing the amplitude of AB oscillations  at all temperatures
down to $T=0$. The crossover between thermal and quantum dephasing
is found to be controlled by the ring perimeter. Our predictions
can be directly tested in future experiments.

\end{abstract}
\end{frontmatter}

\section{Introduction}
Coherent electrons propagating along different paths in multiply
connected conductors, such as, e.g., metallic rings, can interfere
causing a specific quantum contribution to the system conductance
$\delta G$. Threading the ring by an external magnetic flux $\Phi$
one can control the relative phase of the wave functions of
interfering electrons, thus changing the magnitude of $\delta G$
as a function of $\Phi$. The dependence $\delta G (\Phi )$ turns
out to be periodic with the fundamental period equal to the flux
quantum $\Phi_0=hc/e$. These Aharonov-Bohm (AB) conductance
oscillations represent one of the fundamental low temperature
properties of meso- and nanoscale conductors \cite{ArSh}.

In diffusive conductors electrons can propagate along numerous
different paths picking up different phases. Averaging over such
random phases usually washes out AB oscillations $\delta G (\Phi
)$ with the period $\Phi_0$ in the presence of disorder
\cite{ArSh}. There exists, however, a special class of electron
trajectories which interference is not sensitive to averaging over
disorder. These are pairs of time-reversed paths which are also
responsible for the phenomenon of weak localization \cite{CS}. In
disordered rings interference between these trajectories gives
rise to non-vanishing AB oscillations with the principal period
$\Phi_0/2$. Such oscillations will be analyzed below in this
paper.

It is well established that interactions between electrons and
other degrees of freedom can lead to their decoherence thus
reducing electron's ability to interfere. Hence, AB oscillations
can be used as a tool to probe the fundamental effect of
interactions on quantum coherence of electrons in nanoscale
conductors. Recently it was demonstrated \cite{GZ06,GZ08,GZ07}
that the effect of quantum decoherence by electron-electron
interactions can be conveniently studied employing the model of a
system of coupled quantum dots. This model embraces practically
all types of disordered conductors and allows for a
straightforward non-perturbative treatment of electron-electron
interactions. Very recently we employed a similar model in order
to study the effect of electron-electron interactions on AB
oscillations in nanorings with two quantum dots \cite{SGZ}. In
this paper we further extend the approach \cite{SGZ} to nanorings
containing arbitrary number of quantum dots $N$. In the limit of
large $N$ this system serves as a model for diffusive nanorings.

The structure of our paper is as follows. In Sec. 2 we will
address nanorings with two quantum dots \cite{SGZ}. For this
simpler example we will specify our general real time path
integral formalism and recapitulate our main results \cite{SGZ}.
In Sec. 3 we will generalize our analysis adopting it to nanorings
consisting of many quantum dots. The paper is concluded by a brief
discussion in Sec. 4.

\section{Nanorings with two quantum dots}

\subsection{The model and basic formalism}

In this section we will consider the system depicted in Fig. 1.
The structure consists of two chaotic quantum dots (L and R)
characterized by mean level spacing $\delta_L$ and $\delta_R$
which are the lowest energy parameters in our problem. These
(metallic) dots are interconnected via two tunnel junctions J$_1$
and J$_2$ with conductances $G_{t1}$ and $G_{t2}$ forming a
ring-shaped configuration as shown in Fig. 1. The left and right
dots are also connected to the leads (LL and RL) respectively via
the barriers J$_L$ and J$_R$ with conductances $G_L$ and $G_R$. We
also define the corresponding dimensionless conductances of all
four barriers as $g_{t1,2}=G_{t1,2}R_q$ and $g_{L,R}=G_{t1,2}R_q$,
where $R_q=2\pi /e^2$ is the quantum resistance unit.

Following \cite{SGZ} we will assume that dimensionless
conductances $g_{L,R}$ are much larger than unity, while the
conductances $g_{t1}$ and $g_{t2}$ are small as compared to those
of the outer barriers, i.e.
\begin{equation}
g_L,g_R\gg 1,g_{t1},g_{t2}. \label{met}
\end{equation}
 The whole structure is pierced by the magnetic
flux $\Phi$ through the hole between two central barriers in such
way that electrons passing from left to right through different
junctions acquire different geometric phases. Applying a voltage
across the system one induces the current which shows AB
oscillations with changing the external flux $\Phi$.

\begin{figure}
 \centering
\includegraphics[width=2.5in]{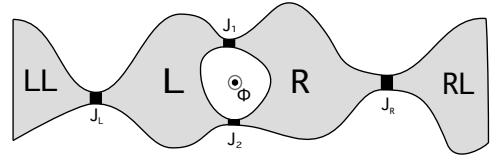}
\caption{\label{f1}The ring-shaped quantum dot structure under
consideration.}
\end{figure}

The system depicted in Fig. 1 is described by the effective
Hamiltonian:
\begin{eqnarray}
\hat H&=&\sum_{i,j=L,R}\frac{C_{ij}\hat {\bm V}_i\hat {\bm
V}_j}{2} +\hat {\bm H}_{LL}+\hat {\bm H}_{RL} \nonumber\\ &&
+\sum_{j=L,R}\hat {\bm H}_{j} +\hat{\bm T}_L +\hat {\bm T}_R+ \hat
{\bm T}, \label{H}
\end{eqnarray}
where $C_{ij}$ is the capacitance matrix, $\hat {\bm V}_{L(R)}$ is
the electric potential operator on the left (right) quantum dot,
$$
\hat {\bm
H}_{LL}=\sum\limits_{\alpha=\uparrow,\downarrow}\int\limits_{LL}d^3{\bm
r} \hat\Psi^\dagger_{\alpha,LL}({\bm r})(\hat
H_{LL}-eV_{LL})\hat\Psi_{\alpha,LL}({\bm r}),
 $$
$$
\hat {\bm
H}_{RL}=\sum\limits_{\alpha=\uparrow,\downarrow}\int\limits_{RL}d^3{\bm
r} \hat\Psi^\dagger_{\alpha,RL}({\bm r})(\hat
H_{RL}-eV_{RL})\hat\Psi_{\alpha,RL}({\bm r})
 $$
are the Hamiltonians of the left and right
leads, $V_{LL,RL}$ are the electric potentials of the leads fixed
by the external voltage source,
$$
\hat {\bm
H}_{j}=\sum\limits_{\alpha=\uparrow,\downarrow}\int\limits_{j}d^3{\bm
r} \hat\Psi^\dagger_{\alpha,j}({\bm r})(\hat H_{j}-e\hat {\bm
V}_{j})\hat\Psi_{\alpha,j}({\bm r})
 $$
defines the Hamiltonians of the left ($j=L$) and
right ($j=R$) quantum dots and
$$\hat H_j=\frac{(\hat p_\mu-\frac e c A_\mu(r))^2}{2m}-\mu+U_j(r)$$
is the one-particle Hamiltonian of electron in $j$-th quantum dot
with disorder potential $U_j(r)$. Electron transfer between the
left and the right quantum dots will be described by the
Hamiltonian
$$
\hat{\bm T}=\sum_{\alpha=\uparrow,\downarrow}\int_{J_1+J_2} d^2{\bm
r}\, \big[t({\bm r})\hat\Psi^\dagger_{\alpha,L}({\bm r})
\hat\Psi_{\alpha,R}({\bm r})+{\rm c.c.}\big].
$$
The Hamiltonian
$\hat {\bm T}_{L(R)}$ describing electron transfer between the
left dot and the left lead (the right dot and the right lead) is
defined analogously.

Following \cite{SGZ} we will describe the time evolution of the
density matrix of our system by means of the standard equation
\begin{equation}
\hat \rho(t)=e^{-i\hat Ht}\hat\rho_0\,e^{i\hat Ht},
\end{equation}
where $\hat H$ is given by Eq. (\ref{H}). Let us express the
operators $e^{-i\hat Ht}$ and $e^{i\hat Ht}$ via path integrals
over the fluctuating electric potentials $V_j^{F,B}$ defined
respectively on the forward and backward parts of the Keldysh
contour:
\begin{eqnarray}
e^{-i\hat Ht}&=&\int  DV_j^F\; {\rm T}\,\exp\left\{-i\int_0^t
dt'\hat H\left[V_j^F(t')\right]\right\},
\nonumber\\
e^{i\hat Ht}&=&\int  DV_j^B\; \tilde{\rm T}\,\exp\left\{i\int_0^t
dt'\hat H\left[V_j^B(t')\right]\right\}.
\end{eqnarray}
Here ${\rm T}\,\exp$ ($\tilde {\rm T}\,\exp$) stands for the time
ordered (anti-ordered) exponent.

Let us define the effective action of our system
\begin{eqnarray}
iS[V^F,V^B]&=&\ln\left( {\rm tr} \left[ {\rm
T}\,\exp\left\{-i\int_0^t dt'\hat H\left[V_j^F(t')\right]\right\}
\right.\right. \nonumber\\ &&\times\, \left.\left. \hat\rho_0
\tilde{\rm T}\,\exp\left\{i\int_0^t dt'\hat
H\left[V_j^B(t')\right]\right\} \right]\right)
\end{eqnarray}
Integrating out the fermionic variables we rewrite the action in
the form
\begin{equation}
    iS=iS_C+iS_{ext}+2 \bf Tr\ln \left[\check  G^{-1} \right].
\label{ac1}
\end{equation}
Here $S_C$ is the standard term describing charging effects,
$S_{ext}$ accounts for an external circuit and
\begin{equation}
  {\bf \check G^{-1}}=\left(\begin{array}{cccc}
   \hat G^{-1}_{LL} & \hat T_L  & 0 & 0 \\
     \hat T^\dag_L & \hat G^{-1}_L & \hat T & 0 \\
     0 & \hat T^\dag & \hat G^{-1}_R & \hat T_R \\
     0 & 0 & \hat T^\dag_R & \hat G^{-1}_{RL}
\end{array}\right).
\end{equation}
is the inverse Green-Keldysh function of electrons propagating in
the fluctuating fields. Here each quantum dot as well as two leads
is represented by the 2x2 matrix in the Keldysh space:
\begin{equation}
   \hat G^{-1}_i=\left(\begin{array}{cc}
    i\partial_t-\hat H_i+eV^F_i & 0 \\
     0 & -i\partial_t+\hat H_i -eV^B_i
\end{array}
\right)
\end{equation}

\subsection{Effective action}

Let us expand the exact action $iS$ (\ref{ac1}) in powers of $\hat
T$. Keeping the terms up to the fourth order in the tunneling
amplitude, we obtain
 \begin{eqnarray}
     iS\approx iS_C+iS_{ext}
     +iS_L+iS_R-2{\bf tr}\left[\hat G_L\hat T \hat G_R\hat T^\dag\right]\nonumber\\
     -{\bf tr}\left[\hat G_L\hat T \hat G_R\hat T^\dag\hat G_L\hat T \hat G_R\hat
     T^\dag\right].
     \label{action}
 \end{eqnarray}
Here $iS_{L,R}$ are the contributions of isolated dots, the terms
$\propto t^2$ yield the Ambegaokar-Eckern-Sch\"on (AES) action
\cite{SZ} $iS^{AES}$ described by the diagram in Fig. 2a, and the
fourth order terms $\propto t^4$ account for the weak localization
correction to the system conductance \cite{GZ08,GZ07}.

\begin{figure}
 \centering
\includegraphics[width=2.5in]{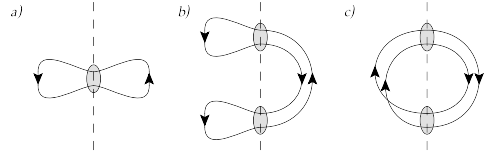}
\caption{\label{f2}Diagrammatic representation of different
contributions originating from expansion of the effective action
in powers of the central barrier transmissions: second order (AES)
terms (a) and different fourth order terms (b,c).}
\end{figure}

It is easy to demonstrate \cite{SGZ} that after disorder averaging
$iS^{AES}$ becomes independent of $\Phi$ and, hence, it does not
account for the AB effect investigated here.  Averaging the last
term in Eq. (\ref{action}) over realizations of transmission
amplitudes and over disorder one can show \cite{SGZ} that only the
contribution generated by the diagram (c) depends on the magnetic
flux. It yields \cite{SGZ}
\begin{eqnarray}
iS^{WL}_{\Phi}=-\frac{ig_{t1}g_{t2}}{4\pi^2 N_L
N_R}\sum\limits_{m,n=1,2}e^{2i(\varphi_g^{(n)}-\varphi_g^{(m)})}\nonumber\\ \times\int
 d\tau_1 d\tau_2
\int dt_1 ...dt_4 C_L(\tau_1)C_R(\tau_2)\qquad\qquad\nonumber\\\times
e^{i(\varphi^+(t_2)-\varphi^+(t_3)+\varphi^+(t_4)-\varphi^+(t_1))}
\sin\frac{\varphi^-(t_1)}{2}\nonumber
\\ \times
\left[h(t_1-t_2-\tau_1)e^{i\frac{\varphi^-(t_2)}{2}}+\right.\qquad\qquad\qquad\qquad\nonumber\\\left.+f(t_1-t_2-\tau_1)e^{-i\frac{\varphi^-(t_2)}{2}}\right]\nonumber\\
\times
\left[h(t_2-t_3-\tau_2)e^{-i\frac{\varphi^-(t_3)}{2}}f(t_3-t_4+\tau_1)-\right.
\quad\nonumber\\ \left.-f(t_2-t_3-\tau_2)e^{i\frac{\varphi^-(t_3)}{2}}h(t_3-t_4+\tau_1)    \right]\nonumber\\
\times \left[e^{i\frac{\varphi^-(t_4)}{2}}f(t_4-t_1+\tau_2)+\right.\qquad\qquad\qquad\qquad\nonumber\\\left.+e^{-i\frac{\varphi^-(t_4)}{2}}h(t_4-t_1+\tau_2)\right]\nonumber\\
+\{L\leftrightarrow R,\varphi^{\pm}\rightarrow -\varphi^{\pm}\},
\label{sw2}
\end{eqnarray}
where $C_{L,R}(t)$ the Cooperons in the left and right dots,
$f(t)=\int f(E)dE/2\pi$ is the Fourier transform of the Fermi
function $f(E)=(\exp (E/T)+1)^{-1}$ and $h(t)=\delta(t)-f(t)$.
Here we also introduced the geometric phases
$\varphi_{g}^{(1,2)}=\frac{e}{c}\int\limits_{L}^{R} dx_\mu
  A_\mu(x)$,
where the integration contour starts in the left dot, crosses the
first ($\varphi_{g}^{(1)}$) or the second ($\varphi_{g}^{(2)}$)
junction and ends in the right dot. The difference between these
two geometric phases is $\varphi_g^{(1)}-\varphi_g^{(2)}=2\pi \Phi
/\Phi_0$. In addition, we defined the ``classical'' and the
``quantum'' components of the fluctuating phase:
$\varphi^+(t)=(\varphi_F(t)+\varphi_B(t))/2$,
$\varphi^-(t)=\varphi_F(t)-\varphi_B(t)$ where the phases $
\varphi_{F,B}(t)=e\int^t d\tau (V^{F,B}_R(\tau)-V^{F,B}_L(\tau))$
are defined on the forward and backward parts of the Keldysh
contour.

The above expression for the action $S^{WL}_{\Phi}$ (\ref{sw2})
fully accounts for coherent oscillations of the system conductance
in the lowest non-vanishing order in tunneling.

\subsection{Aharonov-Bohm conductance}

Let us now evaluate the current $I$ through our system. This
current can be split into two parts, $I=I_0+\delta I$, where $I_0$
is the flux-independent contribution and $\delta I$ is the quantum
correction to the current sensitive to the magnetic flux $\Phi$.
This correction is determined by the action $iS^{WL}_{\Phi}$, i.e.
\begin{equation}
 \delta I=-e\int\mathcal D^2\varphi^{\pm} \frac{\delta S^{WL}_{\Phi}[\varphi^+,\varphi^-]}{\delta \varphi^-(t)}
 e^{iS[\varphi^+,\varphi^-]}.
\label{IAB}
\end{equation}
Below we will only be interested in finding the quantum correction
(\ref{IAB}).

In order to evaluate the path integral over the phases
$\varphi^{\pm}$ in (\ref{IAB}) we note that in the interesting for
us metallic limit (\ref{met}) phase fluctuations can be considered
small down to exponentially low energies \cite{PZ91,Naz99} in
which case it suffices to expand both contributions  up to the
second order $\varphi^{\pm}$. Moreover, this Gaussian
approximation becomes {\it exact} \cite{GGZ05,GZ01,BN} in the
limit of fully open left and right barriers with $g_{L,R} \gg 1$.
Thus, in the metallic limit (\ref{met}) the integral (\ref{IAB})
remains Gaussian at all relevant energies and can easily be
performed.

This task can be accomplished with the aid of the following
correlation functions
\begin{equation}
   \langle\varphi^+(t)\rangle=eVt,\qquad
   \langle\varphi^-(t)\rangle=0,
\label{cf1}
\end{equation}
\begin{equation}
   \langle (\varphi^+(t)-\varphi^+(0))\varphi^+(0)\rangle=-F(t),
\label{cf2}
\end{equation}
\begin{equation}
   \langle \varphi^+(t)\varphi^-(0)+\varphi^-(t)\varphi^+(0)\rangle =2iK(|t|),
\label{cf3}
\end{equation}
\begin{equation}
   \langle \varphi^+(t)\varphi^-(0)-\varphi^-(t)\varphi^+(0)\rangle=2iK(t),
\label{cf4}
\end{equation}
\begin{equation}
   \langle \varphi^-(t)\varphi^-(0)\rangle=0,
\label{cf5}
\end{equation}
where the last relation follows directly from the causality
principle \cite{GZ98}. Here and below we define $V=V_{RL}-V_{LL}$
to be the transport voltage across our system.

Note that the above correlation functions are well familiar from
the so-called $P(E)$-theory\cite{SZ,IN} describing electron
tunneling in the presence of an external environment which can
also mimic electron-electron interactions in metallic conductors.
They are expressed in terms of an effective impedance $Z(\omega)$
``seen'' by the central barriers J$_1$ and J$_2$
\begin{equation}
   F(t)=e^2\int\frac{d\omega}{2\pi}\coth\frac{\omega}{2T}\Re[Z(\omega)]\frac{1-\cos(\omega
   t)}{\omega},
\label{Ft}
\end{equation}
\begin{equation}
  K(t)=e^2\int\frac{d\omega}{2\pi}\Re[Z(\omega)]\frac{\sin(\omega
  t)}{\omega}.
\label{Kt}
\end{equation}
Further evaluation of these correlation functions for our system
is straightforward and yields
\begin{equation}
F(t)\simeq \frac{4}{g} \left(\ln\left|\frac{\sinh(\pi T t)}{\pi
 T\tau_{RC}}\right|+\gamma    \right),
\label{FFF}
\end{equation}
\begin{equation}
K(t)\simeq\frac{2\pi}{g}{\rm sign}(t), \label{KKK}
\end{equation}
where we defined $g=4\pi/e^2Z(0)$ and $\gamma\simeq0.577$ is the
Euler constant. Neglecting the contribution of external leads and
making use of the inequality (\ref{met}) we obtain $g\simeq
2g_Lg_R/(g_L+g_R)$. We observe that while $F(t)$ grows with time
at any temperature including $T=0$, the function $K(t)$ always
remains small and it can be safely ignored in the leading order in
$1/g \ll 1$. After that the Fermi function $f(E)$ drops out from
the final expression for the quantum correction to the current
\cite{GZ08,GZ07,SGZ}. Hence, the amplitude of AB oscillations is
affected by the electron-electron interaction only via the
correlation functions for the ``classical'' component of the
Hubbard-Stratonovich phase $\varphi^+$.

The expression for the current takes the form
\begin{equation}
  \delta I(\Phi )=-I_{AB}\cos(4\pi\Phi/\Phi_0)-I_{WL1}-I_{WL2},
\label{AB+q}
\end{equation}
where the first -- flux dependent -- term in the right-hand side
explicitly accounts for AB oscillations, while the terms
$I_{WL1,2}$ represent the remaining part of the quantum correction
to the current \cite{GZ08} which does not depend on $\Phi$.

Let us restrict our attention to the case of two identical quantum
dots with volume $\mathcal V$, dwell time $\tau_{D}$ and
dimensionless conductances $g_L=g_R \equiv g=4\pi/\delta\tau_D$,
where $\delta=1/ \mathcal V \nu$ is the dot mean level spacing and
$\nu$ is the electron density of states. In this case the
Cooperons take the form $C_L(t;{\bf x},{\bf y})=C_R(t;{\bf x},{\bf
y})=(\theta(t)/\mathcal V)e^{-t/\tau_D}$. We obtain \cite{SGZ}
\begin{eqnarray}
   I_{AB}=\frac{e^2g_{t1}g_{t2}\delta^2V}{4\pi^3}\int\limits_0^\infty d\tau_1 d\tau_2
   e^{-\frac{\tau_1+\tau_2}{\tau_D}-\mathcal F(\tau_1,\tau_2)}.
   \label{res}
\end{eqnarray}
where $\mathcal
F=2F(\tau_1)+2F(\tau_2)-F(\tau_1-\tau_2)-F(\tau_1+\tau_2)$.

In the absence of electron-electron interactions this formula
yields $I_{AB}^{(0)}=4e^2 g_{t1} g_{t2}V/(\pi g^2)$. In order to
account for the effect of interactions we substitute Eq.
(\ref{FFF}) into Eq. (\ref{res}). Performing time integrations at
high enough temperatures we obtain
\begin{equation}
 \frac{ I_{AB}}{I_{AB}^{(0)}}=\left\{\begin{array}{lc}
  e^{-\frac{8\gamma}{g}}\frac{(2\pi T\tau_{RC})^{8/g}}{1+4\pi T\tau_D/g},\quad &    \tau_D^{-1} \lesssim T \lesssim \tau_{RC}^{-1}, \\
 \frac{1}{2\tau_D}\left(\frac{g\tau_{RC}}{T}\right)^{1/2}, &  \tau_{RC}^{-1} \lesssim
 T,
  \end{array}\right.
\end{equation}
while in the low temperature limit we find
\begin{equation}
 \frac{I_{AB}}{I_{AB}^{(0)}}=e^{-\frac{8\gamma}{g}}\left(\frac{2\tau_{RC}}{\tau_D}\right)^{8/g}, \qquad T\lesssim
 \tau_D^{-1}.
 \end{equation}
The above results demonstrate that interaction-induced suppression
of AB oscillations in metallic dots with $\tau_{RC} \ll \tau_D$
persists down to $T=0$. The fundamental reason for this
suppression is that the interaction of an electron with an
effective environment (produced by other electrons) effectively
breaks down the time-reversal symmetry and, hence, causes both
dissipation and dephasing for interacting electrons down to $T=0$
\cite{GZ98}. In this respect it is also important to point out a
deep relation between interaction-induced electron decoherence and
the $P(E)$-theory \cite{SZ,IN} which we already emphasized
elsewhere \cite{GZ08,GZ07}.

\section{Ring composed of a chain of quantum dots}

\begin{figure}[t]
 \centering
\includegraphics[width=2.5in]{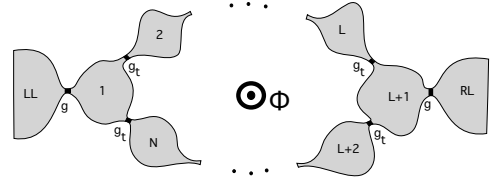}
\caption{\label{f5}Ring composed of $N$ quantum dots}
\end{figure}

Let us now turn to the central part of the present work, i.e. to
the analysis of AB oscillations in nanorings composed of a chain
of quantum dots, as shown in Fig. \ref{f5}. In the previous
section we already demonstrated that the dominant effect of
electron-electron interactions is electron dephasing fully
determined by fluctuations of the phase $\varphi^+$. At the same
time fluctuations of the phase $\varphi^-$ turn out to be
essentially irrelevant for the whole issue. This conclusion is
general being independent of the number of quantum dots in the
ring. Hence, in order address the problem in the many-dot
configuration of Fig. \ref{f5} it suffices to ignore the
fluctuating field $\varphi^-$ and account only for the phase
$\varphi^+$. This observation yields significant simplifications
in our calculation to be presented below.
For simplicity we will consider the case of identical quantum dots
(with mean level spacing $\delta$ and dwell time
$\tau_D=2\pi/(g\delta)$) coupled by junctions with conductances
$g_t$ and the Fano-factor $\beta_t$. Leads are coupled to the ring
at the dots with numbers $1$ and $L+1$ by junctions with
conductance $g$. Interference correction to the conductance of
n-th junction was derived by means of the non-linear sigma-model
approach \cite{GZ06} which yields
\begin{eqnarray}
   \delta G_1=-\frac{e^2  g_t\delta}{4\pi^2}\int\limits_0^\infty dt
   [\beta_t C_{n,n+1}(t)e^{\frac{4\pi i\Phi }{N\Phi_0}}+\nonumber\\+(1-\beta_t)(C_{n,n}(t)+C_{n+1,n+1}(t))+\nonumber\\
   +\beta_tC_{n+1,n}(t)e^{-\frac{4\pi i\Phi }{N\Phi_0}}],
\label{dG1}
\end{eqnarray}
where $C_{m,n}(t)$ is the Cooperon. The
quantum correction to conductance of the whole system can be
obtained with the aid of the Kirchhoff's law. For the case $Ng\ll
g_t$ considered here one finds
\begin{equation}
\delta G=\frac{NL(N-L)g^2}{(2Ng_t+L(N-L)g)^2}\delta g\approx
\frac{L(N-L)g^2}{4Ng_t^2}\delta G_1.
\label{dG}
\end{equation}
In the absence of electron-electron interactions $C_{m,n}(t)$
satisfies the diffusion-like equation which reads
\begin{eqnarray}
    \frac{\partial C_{n,m}(t)}{\partial t}+\frac {2C_{n,m}(t)-C_{n+1,m}(t)e^{-\frac{4\pi i\Phi }{N\Phi_0}}}{2\tau_D}-\nonumber\\-\frac{C_{n-1,m}(t)e^{\frac{4\pi i\Phi }{N\Phi_0}}}{2\tau_D}=\delta_{n,m}\delta(t)
\end{eqnarray}
in the case $n\neq 1,L+1$ and
\begin{eqnarray}
    \frac{\partial C_{n,m}(t)}{\partial t}+\frac {2C_{n,m}(t)-C_{n+1,m}(t)e^{-\frac{4\pi i\Phi }{N\Phi_0}}}{2\tau_D}-\nonumber\\-\frac{C_{n-1,m}(t)e^{\frac{4\pi i\Phi }{N\Phi_0}}}{2\tau_D}+\frac{g\delta}{4\pi}C_{n,m}(t)=\delta_{n,m}\delta(t)
\end{eqnarray}
for $n=1$ or $n=L+1$. The solution of the above diffusion equation
can be represented in the form of the ``functional integral'',
which has the following form:
\begin{eqnarray}
C_{n,m}^{(0)}(t)=\sum\limits_{k=|n-m|}^\infty\sum\limits_{\nu(0)=n}^{\nu(t)=m}
e^{\frac{4\pi i(n-m+NW[\nu(t)])\Phi
}{N\Phi_0}}\times\nonumber\\\times
\int\limits_0^tdt_k\int\limits_0^{t_k}dt_{k-1}...\int\limits_0^{t_2}
dt_1\frac{e^{-\frac{t}{\tau_D}}}{(2\tau_D)^k}.
\end{eqnarray}
Here the summation is performed over all discrete trajectories
with fixed endpoints and $W[\nu(t)]$ denotes the winding number
for a given trajectory.

Let us now include electron-electron interactions. Taking into
account only the $V^+$-component of the fluctuating field one can
easily incorporate the effect of interactions into the above
expression for the Cooperon. One finds
\begin{eqnarray}
C_{n,m}(t)=\sum\limits_{k=|n-m|}^\infty\sum\limits_{\nu(0)=n}^{\nu(t)=m}
e^{\frac{4\pi i(n-m+NW[\nu(t)])\Phi }{N\Phi_0}}
\times\nonumber\\\times\int\limits_0^tdt_k...\int\limits_0^{t_2}
dt_1 \frac{e^{-\frac{t}{\tau_D}+ie\int\limits_0^t d\tau
(V^+_{\nu(\tau)}(\tau)-V^+_{\nu(\tau)}(t-\tau))}} {(2\tau_D)^k},
\label{cmn}
\end{eqnarray}
i.e. the fluctuating field $V^+$ just modifies the phases of the
electron wave functions. Averaging over Gaussian fluctuations of
$V^+$ we get
\begin{eqnarray}
\Bigl\langle \exp\Bigl[ie\int\limits_0^t d\tau (V_{\nu(\tau)}(\tau)-V_{\nu(\tau)}(t-\tau))\Bigl]\Bigr\rangle_{V^+}=\nonumber\\
=\exp\Bigl[-e^2\int\limits_0^t d\tau_1 d\tau_2
(F_{\nu(\tau_1),\nu(\tau_2)}(\tau_1-\tau_2)-\nonumber\\-F_{\nu(\tau_1),\nu(\tau_2)}(t-\tau_1-\tau_2))\Bigr].
\label{41}
\end{eqnarray} Here $F_{m,n}(t)=\langle V^+_m(t)
V^+_n(0)\rangle_{V^+}$ defines the correlator for fluctuating
voltages.

In order to evaluate the Cooperon in the presence of interactions
let us first expand the exponent in Eq. (\ref{41}) in Taylor
series, then perform the summation over all trajectories and after
that re-exponentiate the result. This procedure is equivalent to
the substitution $\langle\langle e^F\rangle\rangle\rightarrow
e^{\langle\langle F\rangle\rangle}$ which -- although not exact --
is known to provide sufficiently accurate results for the problem
in question at all time scales (cf., e.g., Ref. \cite{TM}).

Averaging over diffusive pathes is performed with the aid of the
diffuson $D_{m,n}(t)$:
\begin{eqnarray}
\langle\langle F_{\nu(\tau_1),\nu(\tau_2)}(\tau_1-\tau_2)\rangle\rangle=\qquad\qquad\qquad\qquad\nonumber\\=\frac{1}{N}\sum\limits_{m,n=1}^N F_{m,n}(\tau_1-\tau_2)D_{m,n}(|\tau_1-\tau_2|)
\end{eqnarray}
As a result one finds \cite{GZ07}
\begin{equation}
  C_{m,n}(t)=C_{m,n}^{(0)}(t)e^{-\mathcal F(t)},
\end{equation}
where
\begin{eqnarray}
\mathcal F(t)=\frac{e^2}{N}\sum\limits_{n,m=1}^N\int\limits_0^t
d\tau_1 d\tau_2 F_{m,n}(\tau_1-\tau_2)\times\nonumber\\
\times\left(D_{m,n}(|\tau_1-\tau_2|)-
D_{m,n}(|t-\tau_1-\tau_2|)\right).
\end{eqnarray}
The correlator for fluctuating voltages can be derived, e. g., by
means of the non-linear sigma model \cite{GZ06} which yields
\begin{equation}
 F_{m,n}(t)=\frac{\tau_D}{N}\sum\limits_{q=1}^N\int\frac{d\omega}{2\pi}e^{-i\omega t}\omega\coth\frac{\omega}{2T}\frac{f(q)e^{\frac{2\pi i q}{N}(m-n)}}{\omega^2\tau_D^2+\varepsilon^2(q)}
\end{equation}
where \begin{equation}
 f(q)=\frac{g_t\tau_D
 e^2}{\pi}\frac{\epsilon(q)}{(4C\epsilon(q)+C_g)^2},
\end{equation}
\begin{equation}
\varepsilon(q)=\epsilon(q)+\frac{g_t\tau_D e^2}{\pi}\frac{\epsilon(q)}{4C\epsilon(q)+C_g}
\end{equation}
and $\epsilon(q)=1-\cos\frac{2\pi q}{N}$. As above, here $C$ and
$C_g$ denote respectively the junction and the dot capacitances.

Finally we specify the expressions for the diffuson and the
Cooperon in the absence of electron-electron interactions. They
read
\begin{equation}
  D_{m,n}(t)=\frac{\tau_D}{N}\sum\limits_{q=1}^N\int\frac{d\omega}{2\pi}
  \frac{e^{-i\omega t+\frac{2\pi i
  q}{N}(m-n)}}{-i\omega\tau_D+\epsilon(q)}.
\end{equation}
\begin{equation}
  C_{m,n}^{(0)}(t)=\frac{\tau_D}{N}\sum\limits_{q=1}^N
  \int\frac{d\omega}{2\pi}\frac{e^{-i\omega t+\frac{2\pi i
  q}{N}(m-n)}}{-i\omega\tau_D+\epsilon(q-2\Phi/\Phi_0)}.
\end{equation}

The above equations are sufficient to evaluate the function
$\mathcal F(t)$ in a general form. Here we are primarily
interested in AB oscillations and, hence, we only need to account
for the flux-dependent contributions determined by the electron
trajectories which fully encircle the ring at least once.
Obviously, one such traverse around the ring takes time $t\geq
N^2\tau_D$. Hence, the behavior of the function $\mathcal F(t)$
only at such time scales needs to be studied for our present
purposes. In this long time limit $\mathcal F(t)$ is a linear
function of time with the corresponding slope
\begin{eqnarray}
 \mathcal F'(t\geq N^2\tau_D)\approx\qquad\qquad\qquad\qquad\qquad\qquad\qquad\nonumber\\\approx\frac{2e^2\tau_D^2}{N}\sum\limits_{q=1}^{N-1}\int\frac{d\omega}{2\pi}\frac{f(q)\epsilon(q)\omega\coth\frac{\omega}{2T}}{(\omega^2\tau_D^2+\epsilon^2(q))(\omega^2\tau_D^2+\varepsilon^2(q))}
 \label{def}
\end{eqnarray}
This observation implies that at such time scales
electron-electron interactions yield exponential decay of the
Cooperon in time
\begin{equation}
C_{m,n}(t)\approx
C_{m,n}^{(0)}(t)e^{-\frac{t}{\tau_\phi}}
\end{equation}
where
\begin{equation}
\frac{1}{\tau_\phi}=\mathcal F'(t\geq N^2\tau_D)
\label{dephtime}
\end{equation}
is the effective dephasing time for our problem. In the case
$C_g\gg C$ and $\tau_D\gg\tau_{RC}\equiv 2\pi C_g/(e^2 g_t)$ from
Eq. (\ref{dephtime}) we obtain
 \begin{equation}
    \frac{1}{\tau_\phi}=\left\{\begin{array}{cc}
    \frac{\delta}{\pi}\ln\frac{4E_C}{\delta} & \qquad\qquad T\ll 1/N\tau_D ,\\
       \frac{\pi N T }{3 g_t} & \qquad\qquad T\gg 1/N\tau_D,
    \end{array} \right.
\label{dephtime}
\end{equation} where $E_C=e^2/(2C_g)$. These
expressions are fully consistent with recent results
\cite{GZ08,GZ07} derived for chains of quantum dots (or
scatterers). It is also important to emphasize that in the case of
weakly disordered diffusive conductors the expression for
$\tau_\phi$ (\ref{dephtime}) in the limit of low $T$ coincides
with that obtained earlier within different theoretical approaches
\cite{GZ98,GZ03}. For further discussion of this point we refer
the reader to Ref. \cite{GZ07}.

Let us emphasize again that the above results for $\mathcal F(t)$
apply at sufficiently long times which is appropriate in the case
of AB conductance oscillations. At the same time, other physical
quantities, such as, e.g., weak localization correction to
conductance can be determined by the function $\mathcal F(t)$ at
shorter time scales. Our general results allow to easily recover
the corresponding behavior as well. For instance, at $T\gg\tau_D$
and $t \ll N^2\tau_D$ we get
\begin{equation}
   \mathcal F(t)\approx \frac{4T }{3 g_t}\left(\frac{2\pi}{\tau_D}\right)^{1/2} t^{3/2}+...
\end{equation}
in agreement with the results \cite{GZ07}. This expression yields
the well known dependence $\tau_\phi \propto T^{-2/3}$ which -- in
contrast to Eq. (\ref{dephtime}) -- does not depend on $N$ and
remains applicable in the high temperature limit.

To proceed further let us integrate the expression for the
Cooperon over time. We obtain
\begin{eqnarray}
  \int\limits_0^{\infty}C_{m,n}(t)dt=\qquad\qquad\qquad\qquad\qquad\qquad\nonumber\\
  =\frac{\tau_D}{N}\sum\limits_{q=1}^N\frac{e^{\frac{2\pi i q}{N}(m-n)}}{\epsilon(q-2\Phi/\Phi_0)+\tau_D/\tau_{\phi}+g/(g_t
  N)},
\label{tint}
\end{eqnarray}
where the term $g/(g_t N)$ in the denominator accounts for the
effect of external leads and remains applicable as long as $Ng\ll
g_t$. Combining Eqs. (\ref{dG1}), (\ref{dG}) and (\ref{tint})
after summation over $q$ we arrive at the final result
\begin{eqnarray}
\delta G^{AB}=\frac{e^2L(N-L)g^2}{2\pi Ng_t^2}\qquad\qquad\qquad\qquad\nonumber\\
\times\frac{(\beta_t\alpha+1-\beta_t)(z^{-N}-\cos(4\pi\Phi/\Phi_0))}{\sqrt{\alpha^2-1}(z^N+z^{-N}-2\cos(4\pi\Phi/\Phi_0))},
\label{final}
\end{eqnarray}
where $\alpha=1+\frac{\tau_D}{\tau_\phi}+\frac{g}{g_tN}$ and
$z=\alpha+\sqrt{\alpha^2-1}$.

Eq. (\ref{final}) is the central result of the present paper.
Together with Eq. (\ref{dephtime}) it fully determines AB
oscillations of conductance in nanorings composed of metallic
quantum dots in the presence of electron-electron interactions.

Expanding Eq. (\ref{final}) in Fourier series we obtain
\begin{equation}
   \delta G^{AB}=\sum\limits_{k=1}^{\infty} \delta G^{(k)}\cos\left(4\pi k\Phi/\Phi_0\right)
\end{equation}
where
\begin{equation}
\delta G^{(k)}=-\frac{e^2L(N-L)g^2(\beta_t\alpha+1-\beta_t)}{2\pi N g_t^2\sqrt{\alpha^2-1}}z^{-N|k|}
\end{equation}
In the limit $\tau_\phi\gg\tau_D$ we have $ z\approx
1+\sqrt{2\tau_D/\tau_\phi}+...$, hence $\delta G^{(k)}$ behaves as
\begin{equation}
\delta G^{(k)}\propto e^{-N|k|\sqrt{\frac{2\tau_D}{\tau_\phi}}},
\end{equation}
i.e. at hight temperatures $\log|\delta G|$ scales with $N$ as
$N^{3/2}$ while at low temperatures it scales as $N$. The
temperature dependence of the first three harmonics of AB
conductance in the presence of electron-electron interactions is
depicted in Fig. 4.

\begin{figure}[t]
 \centering
\includegraphics[width=2.5in]{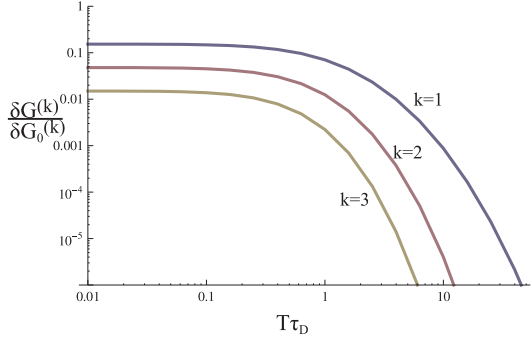}
\caption{\label{f6}Temperature dependence of the first three
harmonics of AB conductance for $g_t=500$, $g=30$, $N=10$,
$\beta_t=1$ and $\tau_D/\tau_{RC}=120$.}
\end{figure}

\section{Discussion}

The results obtained here allow to formulate quantitative
predictions regading the effect of electron-electron interactions
on Aharonov-Bohm oscillations of conductance for a wide class of
disordered nanorings embraced by our model. Of particular interest
is the situation of large number of dots $N \gg 1$ which
essentially mimics the behavior of diffusive nanostructures. In
order to establish a direct relation to this important case it is
instructive to introduce the diffusion coefficient
$D=d^2/(2\tau_D)$ and define the electron density of states
$\nu=1/(d^3\delta)$, where $d$ is a linear dot size.  Then we
obtain with exponential accuracy:
\begin{equation}
\delta G^{(k)}\sim\left\{ \begin{array}{lc}
e^{-|k| (\mathcal L/\mathcal L_\phi)} & \qquad\qquad T\ll D/(\mathcal L d),\\
e^{-|k| (\mathcal L/\mathcal L_\phi)^{3/2}} & \qquad\qquad T\gg
D/(\mathcal Ld).
\end{array}
\right.
\nonumber
\end{equation}
Here we introduced the ring perimeter $\mathcal L=Nd$ and the
effective decoherence length
\begin{equation}
\mathcal L_\phi=\left\{ \begin{array}{lc}
 \left(\frac{\pi\nu d^3 D}{\ln\frac{4E_C}{\delta}}\right)^{1/2}& \qquad\qquad T\ll D/(\mathcal L d),\\
\left(\frac{12\nu d^2 D^2}{T}\right)^{1/3} & \qquad\qquad T\gg
D/(\mathcal L d).
\end{array}
\right. \nonumber
\end{equation}
Note in the high temperature limit $T\gg D/(\mathcal L d)$ the
above results match with those derived earlier for metallic
nanorings with the aid of different approaches \cite{TM,LM}. On
the other hand, at lower $T$ our results are different. This
difference is due to low temperature saturation of $\tau_\phi$
which was not accounted for in Refs. \cite{TM,LM}. A non-trivial
feature predicted here is that -- in contrast to weak localization
\cite{GZ98} -- the crossover from thermal to quantum dephasing is
controlled by the ring perimeter $\mathcal L$. This is because
only sufficiently long electron paths fully encircling the ring
are sensitive to the magnetic flux and may contribute to AB
oscillations of conductance.

We believe that the quantum dot rings considered here can be
directly used for further experimental investigations of quantum
coherence of interacting electrons in nanoscale conductors at low
temperatures.

\section*{ Acknowledgments}

\vspace{0.3cm}

We would like to thank D.S. Golubev for numerous illuminating
discussions. This work was supported in part by RFBR grant
09-02-00886. A.G.S. also acknowledges support from the Landau
Foundation and from the Dynasty Foundation.


\begin{thebibliography}{30}
\bibitem{ArSh} A.G. Aronov, Yu.V. Sharvin,
Rev. Mod. Phys. 59 (1987) 755.
\bibitem{CS} S. Chakravarty, A. Schmid,
Phys. Rep. 140 (1986) 193.
\bibitem{GZ06} D.S. Golubev, A.D. Zaikin, Phys. Rev. B 74 (2006) 245329.
\bibitem{GZ08} D.S. Golubev, A.D. Zaikin, New J. Phys. 10 (2008) 063027.
\bibitem{GZ07} D.S. Golubev, A.D. Zaikin, Physica E 40 (2007) 32.
 \bibitem{SGZ} A.G. Semenov, D.S. Golubev, A.D. Zaikin, Phys. Rev. B 79 (2009) 115302.
\bibitem{SZ} G. Sch\"on, A.D. Zaikin, Phys. Rep. 198 (1990) 237.
\bibitem{PZ91}  S.V. Panyukov, A.D. Zaikin, Phys. Rev. Lett. 67 (1991) 3168.
\bibitem{Naz99} Yu.V. Nazarov,  Phys. Rev. Lett. 82 (1999) 1245.
\bibitem{GGZ05} D.S. Golubev, A.V. Galaktionov, A.D. Zaikin, Phys. Rev. B
    72 (2005) 205417.
\bibitem{GZ01} D.S. Golubev, A.D. Zaikin, Phys. Rev. Lett. 86 (2001) 4887;
Phys. Rev. B 69 (2004) 075318.
\bibitem{BN} D.A. Bagrets, Yu.V. Nazarov, Phys. Rev. Lett. 94 (2005) 056801.
\bibitem{GZ98} D.S. Golubev, A.D. Zaikin, Phys. Rev. Lett. 81 (1998) 1074;
Phys. Rev. B 59 (1999) 9195; Phys. Rev. B 62 (2000) 14061; Physica
B 255 (1998) 164.
\bibitem{GZ03} D.S. Golubev, A.D. Zaikin, J. Low Temp. Phys. 132
(2003) 11.
\bibitem{IN} G.L. Ingold, Yu.V. Nazarov, {\it Single Charge Tunneling},
 (Plenum Press, New York) {\it NATO ASI Series} B 294 (1992) p. 21.
\bibitem{TM} C. Texier, G. Montambaux, Phys. Rev. B 72 (2005) 115327.
\bibitem{LM} T. Ludwig, A.D. Mirlin, Phys. Rev. B 69 (2004) 193306.
\end{thebibliography}
\end{document}